# Clocked Bursts From GS 1826-238

M. Cocchi[(1)], A. Bazzano[(1)], L. Natalucci[(1)], P. Ubertini[(1)]
J. Heise[(2)], J. J. M. in't Zand[(2)], J. M. Muller[(2,3)], M. J. S. Smith[(2,3)]

*(1) Istituto di Astrofisica Spaziale, C.N.R., Via del Fosso del Cavaliere 00133 Roma- Italy*
*(2) Space Research Organisation of the Netherlands (SRON), Utrecht, the Netherlands*
*(3) also BeppoSAX Science Data Centre, Nuova Telespazio, Roma, Italy*

**ABSTRACT**. We report on the long term monitoring of the GINGA transient source GS 1826-238 performed with the BeppoSAX Wide Field Camera (WFC) instrument, during four different observing campaigns covering October 1996 - September 1999.
WFC detection of type-I bursting activity from the source ruled out its proposed Black Hole candidacy and clearly suggested the compact object related to GS 1826-238 to be a weakly magnetized neutron star. The analysis of the arrival times of the observed 78 bursts lead to the discovery of a recurrence of ~5.75 hours with a spread of 38 minutes (FWHM) along more than 3 years monitoring data [15].
We performed a more detailed analysis of the whole available data, and evidence of shortening of the recurrence time, together with a drastic narrowing in the spread (down to a few minutes) was observed on a one year time scale. Possible relation with the source X-ray persistent emission is discussed.

## INTRODUCTION

GS 1826-238 was serendipitously discovered by GINGA on September 8, 1988 with an average X-ray intensity of about 26 mCrab in the 1-40 keV range [9]. During September 9-16, 1988, rapid fluctuations (flickering) were observed on time scales down to 2 ms with an rms variation of 30% [12]. In the GINGA All Sky Monitor the source was below 50 mCrab all along the period from August through October 1988 [13]. Since no detection was available from previous satellite observations, GS 1826-238 was tentatively reported as transient. Similarity with Cyg X-1 and GX 339-4 in the low (hard) state suggested to consider the source as a black hole candidate [12]. The GINGA spectrum was in fact well fitted by a single power law with photon index ~1.7.

In 1989 the source was detected by TTM on March 17 [6] at a flux of about 32 mCrab (2-28 keV).

Later on the source was observed by the ROSAT PSPC (October 1990, June and October 1992, see [1]) and no X-ray bursts were detected during 8 hours of net exposure time. The spectrum was well fitted by a power law with a photon index in the range 1.5-1.8 and an absorption $N_H = 5 \times 10^{21}$ H cm$^{-2}$. Follow-up optical studies led to the identification of a V=19.3 optical counterpart.

The ROSAT source was inside the GINGA error box and a larger one of an unidentified X-ray burster [16] observed with OSO-8 [4], OSO 7 Catalogue [10], and containing the source 4U1831-23 [5], which is also present in the ARIEL V catalogue [17].

During November 1994 the source was detected with OSSE at 7.5 standard deviations in the 60-200 keV range [11].

The BeppoSAX WFC discovery of type-I bursting activity from GS 1826-238 since August 1996 [3],[15]) solved the question about the source nature, clearly pointing to a neutron star Low Mass Binary (LMXB) scenario. Moreover its very stable X-ray emission over more than 3 years suggested the authors to classify GS 1826-238 as a weak persistent source rather than a transient. The hypothesis of a neutron star LMXB system was previously discussed by Strickman et al. (1996) and Barret et al. (1996).

## OBSERVATION AND DATA ANALYSIS

The Wide Field Cameras (Jager et al., 1997) on board BeppoSAX are designed for performing spatially resolved simultaneous monitoring of X-ray sources in crowded fields enabling studies of spectral variability at high time resolution. The mCrab sensitivity in 2-28 keV over a large (40°×40° deg.) field of view (FOV) and the near-to-continuos operation over a period of years offers for the first time the unique opportunity to measure continuum emission as well as bursting behavior from new as well as already known (weak) transients. For this reason the Galactic Bulge has been monitored over 1 to 2 months during each of the visibility periods since the beginning of the BeppoSAX operational life in July 1996. During those observations, amounting to a total of ~2.1 Ms live time [14], at least 45 sources and about 700 bursts have been detected.

The sky region containing GS1826-238 was monitored since August 20th 1996. Whenever observed with the WFCs the source shows an average 2-28 keV persistent emission of 31 mCrab, varying from 22 to 39 mCrab. This corresponds to an average flux of $1.1 \times 10^{-9}$ erg cm$^{-2}$ s$^{-1}$ in the 2-28 keV range. The long term 2-10 keV monitoring performed with the ASM experiment on board RXTE confirms the WFC results.

An extensive search for X-ray bursts from GS 1826-238 was performed on the whole available data set, thus leading to the identification of 78 events in the period August 1996 - April 1998. A preliminary time analysis of the bursts occurrence time lead to the discovery that the wait time between the events is almost constant, being ~5.75 hours with a spread of only 38 minute FWHM [15]. Regular type-I bursting was already found in other sources, e.g. 1658-298, 1820-303 and 1323-619 [8]. Nevertheless, the stability of the clock for over 3 years is a unique feature of GS1826-238.

The burst recurrence was investigated in more detail by analyzing the 1996, 1997 and 1998 data separately, to search for possible differences in the clock on one year time scale. The results, which are plotted in Figure 1, show some interesting features.

First of all, the spread of the wait times in 1997 and 1998 is much less (only ~5 minutes FWHM) than the one obtained by Ubertini et al. 1999 on the total 3 years data set (~38 minutes FWHM, see Table 1). Besides that, the wait times decreased significantly in 1998 with respect to 1997 and 1996. Despite the narrow distributions in 1997 and 1998, the wait times distribution in 1996 is very broad and, together with the period decrease in 1998, accounts for the much larger FWHM observed on the 3 years time span.

Such regular bursting behavior on a wide time interval imposes tight constraints on the parameters of the burst engine, and so the standard model of type-I bursts needs to be checked to explain the features of this almost unique binary system.

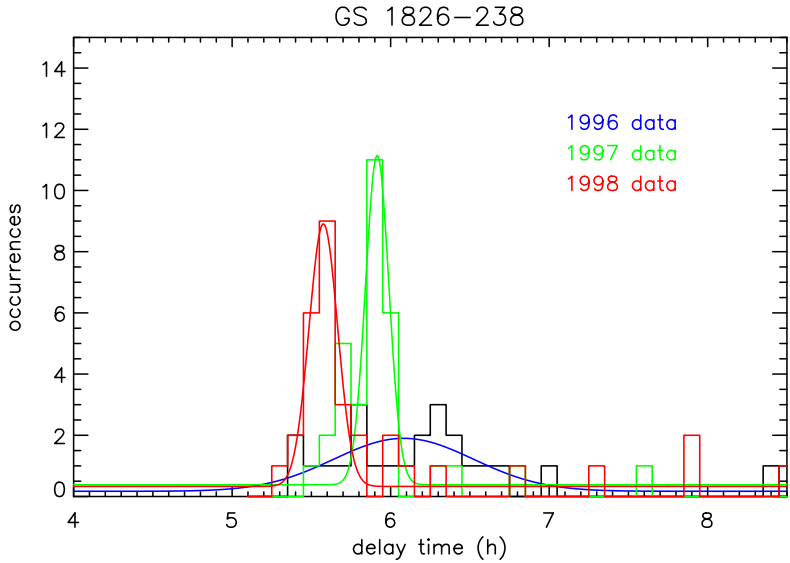

**FIGURE 1.** Burst wait times distribution in 1996,1997,1998.

**Table 1.** Burst wait times and persistent emission distributions in 1996,1997,1998

|  | 1996 | 1997 | 1998 |
|---|---|---|---|
| Recurrence time (h) | 6.08 ± 0.14 | 5.92 ± 0.01 | 5.58 ± 0.01 |
| Dispersion (h, 1 sigma) | 0.44 ± 0.15 | 0.07 ± 0.01 | 0.09 ± 0.01 |
| WFC 2-28 keV average | | | |
| intensity (mCrab) | 30.3 ± 2.2 | 31.4 ± 2.3 | 31.9 ± 2.0 |
| Dispersion (1 sigma) | 5.5 mCrab | 4.5 mCrab | 2.9 mCrab |
| ASM 2-10 keV average | | | |
| intensity (mCrab) | 24.6 ± 0.2 | 28.1 ± 0.3 | 27.8 ± 0.5 |
| Dispersion (1 sigma) | 9.2 mCrab | 12.0 mCrab | 11.9 mCrab |

# DISCUSSION

Despite its X-ray variability, GS 1826-238 is the first example of clocked burst activity for more than 3 years. This, in terms of standard burst model, points to a very stable matter accretion for a long time.

It is a well established statement that the X-ray luminosity is correlated to the accretion rate on the compact object in the binary system. So, under the hypothesis that the burst energetics did not vary, we investigated if slight changes in the source persistent emission influenced the variation of the recurrence period or the spread of the burst wait times distribution. In particular, one could expect increased accretion, and so higher average X-ray intensity, in 1998 when the burst period decreased from 5.92 h to 5.58 h. Moreover, less stable accretion, and so higher X-ray variability, could explain the observed 1996 wait time distribution. A preliminary analysis of the GS 1826-238 bursts fluences supports the assumption made.

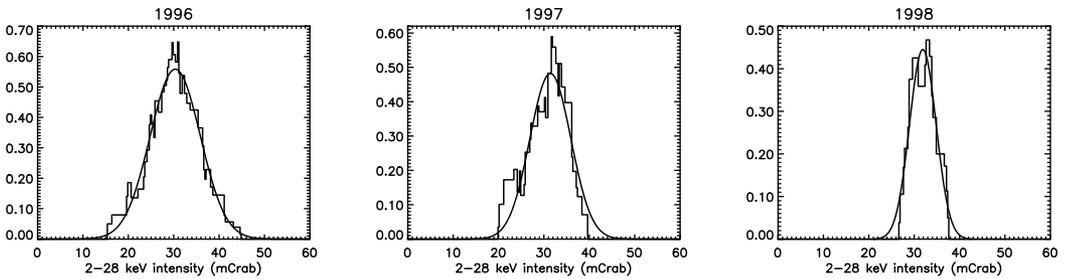

**FIGURE 2.** WFC source intensity distributions in 1996, 1997, 1998

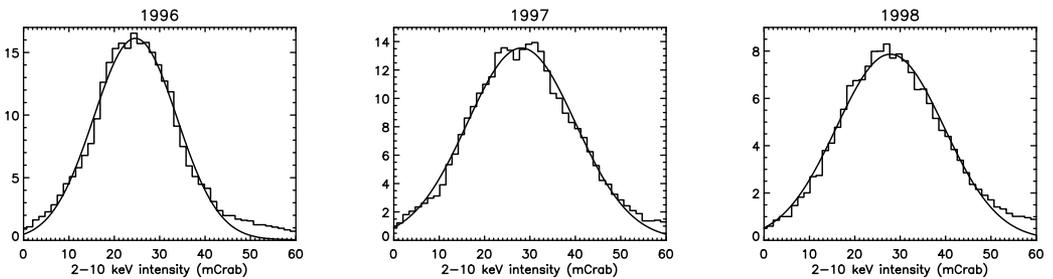

**FIGURE 3.** ASM source intensity distributions in 1996, 1997, 1998

We plotted (Figure 2 and 3) the distribution of the source X-ray intensities measured by BeppoSAX WFC in 2-28 keV and by RXTE-ASM (2-10 keV) in 1996, 1997 and 1998 respectively. We observe no evident correlation between the burst wait times, which varied by ~8% in 1 year, and the source persistent intensity. There is also no evident correlation between the "dispersions" of the wait times and the X-ray intensity, but perhaps only marginal evidence in the 1996

WFC data, possibly implying an energy dependence effect which needs further investigation.

If the wait times are strictly related with the accretion rate on the neutron star, $M_{dot}$ variation is expected in 1997-1998. This is not apparent from the measured persistent X-ray intensity, which is almost constant. One could deduce that the X-ray source intensity is only a rough indicator of the mass transfer rate.

Further analysis is needed to better understand the peculiar features of GS 1826-238. For example, all WFC 1999 data (25% of the total available) need still to be included in the burst wait times analysis. Moreover, more accurate spectral data on a wide energy band of both the bursts and the persistent emission could better explain the source behavior. To this end, a new wide band BeppoSAX NFI observation campaign is planned within October 1999.

## ACKNOWLEDGEMENTS

We also thank Team Members of the BeppoSAX Science Operation Center and Science Data Center for their continuos support and timely actions for quasi "real time" detection of new transient and bursting sources and the follow-up TOO observations.